# Coexistence of magnetism and superconductivity in CeRh$_{1-x}$Ir$_x$In$_5$

P.G. Pagliuso[1], C. Petrovic[1,2], R. Movshovich[1], D. Hall[2], M.F. Hundley[1], J.L. Sarrao[1], J.D. Thompson[1], and Z. Fisk[1,2]

[1]*Los Alamos National Laboratory, Los Alamos, NM 87545*

[2]*National High Magnetic Field Laboratory, Florida State University, Tallahassee, FL 32310*

Version Date: January 19, 2001

Abstract

We report a thermodynamic and transport study of the phase diagram of CeRh$_{1-x}$Ir$_x$In$_5$. Superconductivity is observed over a broad range of doping, $0.3 < x \leq 1$, including a substantial range of concentration ($0.3 < x < 0.6$) over which it coexists with magnetic order (which is observed for $0 \leq x < 0.6$). The anomalous transition to zero resistance that is observed in CeIrIn$_5$ is robust against Rh substitution. In fact, the observed bulk $T_c$ in CeRh$_{0.5}$Ir$_{0.5}$In$_5$ is more than double that of CeIrIn$_5$, whereas the zero-resistance transition temperature is relatively unchanged for $0.5 < x < 1$.

In conventional superconductors magnetism and superconductivity are usually antithetical: an internal magnetic field breaks time-reversal symmetry which kills BCS superconductivity.[1] In a few cases (e.g., $NdRh_4B_4$, $TbMo_6S_8$, $DyNi_2B_2C$), superconductivity and magnetic order coexist, but this rare situation arises from magnetic order among localized 4f electrons that are uncoupled to itinerant conduction electrons which form the superconducting condensate.[2] In contrast, unconventional superconductivity in heavy Fermion superconductors (HFSs), like the high-$T_c$ cuprates, relies on some form of magnetic coupling to produce superconductivity.[3] These HFSs appear to fall into one of two classes: those in which ordered magnetism coexists with superconductivity and those in which ordered magnetism competes with superconductivity. Most U-based HFSs belong to the first category, while the Ce-based superconductors belong to the second.[4] For example, $UPt_3$, $URu_2Si_2$, $UNi_2Al_3$, and $UPd_2Al_3$ develop superconductivity out of an antiferromagnetically ordered state that persists to T=0. On the other hand, the ground state of the prototypical HFS $CeCu_2Si_2$ can be either magnetic or superconducting depending on small (< 1%) variations in stoichiometry or the application of small (<5 kbar) applied pressures.[5] $CeCu_2Ge_2$, $CeRh_2Si_2$, $CePd_2Si_2$, and $CeIn_3$ display more stable antiferromagnetic ground states, but with the application of pressures ranging from 10 to 100 kbar, superconductivity can be induced when magnetism is suppressed.[3] Although small windows of superconductivity can exist before magnetism is completely suppressed in some of these materials, this is generally attributed to 'real-world' effects such as pressure and stoichiometric inhomogeneities.[6]

Here, we report a striking counterexample to the above categorization. $CeRh_{1-x}Ir_xIn_5$ (0.3 < x < 0.6) displays superconductivity, with critical temperature $T_c$ ~ 1 K over a wide range of composition, that develops out of and coexists with a magnetically ordered state, with Neel temperature $T_N$ ~ 4 K. The broad range of composition over which superconductivity is observed (0.3 < x ≤ 1) is also counter to expectation: Small amounts of chemical disorder in either U- or Ce-based HFSs are generally sufficient to destroy heavy-Fermion superconductivity.[7]

$CeRhIn_5$ is a heavy Fermion antiferromagnet ($T_N$ =3.8 K, $\gamma$=400 mJ/molK$^2$, where $\gamma$ is the linear-in-T coefficient of heat capacity C at low temperature) in which superconductivity can be induced with applied pressure (at critical pressure $P_c$=16 kbar, $T_c$=2.1 K, $\gamma$=400 mJ/molK$^2$).[8,9] Unlike the Ce-based materials discussed above, in CeRhIn$_5$ $T_N$ is essentially independent of pressure before disappearing, and the transition to superconductivity appears to be first order. The crystal structure in which CeRhIn$_5$ crystallizes is also host to two ambient-pressure heavy Fermion superconductors: CeIrIn$_5$ ($T_c$=0.4 K, $\gamma$=720 mJ/molK$^2$)[10] and CeCoIn$_5$ ($T_c$=2.3 K, $\gamma$=290 mJ/molK$^2$).[11] CeIrIn$_5$ displays the additional feature that it exhibits a zero-resistance transition at 1.2 K, well above the bulk $T_c$.[10] In order to understand CeRhIn$_5$'s unconventional magnetic behavior and the development of zero-resistance and bulk superconductivity in CeIrIn$_5$, we have performed a detailed study of the series of isovalent alloys CeRh$_{1-x}$Ir$_x$In$_5$. Our principal results, summarized in Fig. 1, are reported below: antiferromagnetism persists for $0 < x < 0.6$ and is lost rather abruptly as a function of Ir concentration; superconductivity is observed over a broad range of doping, $0.3 < x \leq 1$ – i.e., up to 70% of the Ir ions can be replaced and yet superconductivity is retained; and the separation in temperature between the zero resistance feature and bulk superconductivity decreases with increasing Rh doping, approaching zero separation at CeRh$_{0.5}$Ir$_{0.5}$In$_5$.

Single crystals of CeRh$_{1-x}$Ir$_x$In$_5$ with characteristic size 1 cm$^3$ were grown from an In flux.[8] Room-temperature x-ray powder diffraction measurements on crushed single crystals revealed that the materials were single phase and crystallized in the primitive tetragonal HoCoGa$_5$ structure.[12] In this structure, CeRh$_{1-x}$Ir$_x$In$_5$ can be viewed as layers of CeIn$_3$ stacked sequentially along the c-axis with intervening layers of "Rh$_{1-x}$Ir$_x$In$_2$." No superlattice peaks at e.g., x=1/2 were observed. Lattice constants of CeRh$_{1-x}$Ir$_x$In$_5$ extracted from these measurements are shown in Fig. 2. The *a* lattice constant, which is the nearest-neighbor Ce-Ce spacing in this structure, expands with increasing x, while the *c*-axis shrinks.[12] To the extent that the measured lattice constants follow Vegard's law, taking x as the nominal Rh:Ir ratio of the starting constituents, the nominal composition agrees well with the actual composition. From these data, we estimate an uncertainty in x of +/- 0.05 at a given composition, a variation consistent with independent estimates from analysis of high-temperature magnetic susceptibility data and in crystal-to-crystal variations in ground-state properties. However, the value of x in a given crystal is always well defined; we observe no evidence for concentration inhomogeneity or phase segregation as judged by the sharpness of diffraction peaks and the low values for residual resistivity ($\rho(T\rightarrow 0) < 10$ $\mu\Omega$cm for all x) that are observed across the series, as well as the sharpness of the phase transitions observed in a given crystal.

In Fig. 3 we show C/T and magnetic entropy S as a function of T for representative x. Although the ground state changes from antiferromagnetic (x < 0.3) to superconducting coexisting with antiferromagnetism (0.3 < x < 0.6) to superconducting (x > 0.6), the total magnetic entropy evolved by 6 K is nearly independent of x. This indicates that the same heavy electrons are producing the variety of observed ground states and, in particular, the coexisting magnetic order and superconductivity where it is observed. In the following, we discuss each of these ranges in greater detail.

For low x (x < 0.3), a single phase transition is observed in heat capacity whose shape and character is similar to that of stoichiometric CeRhIn$_5$, in which antiferromagnetic order with Q=(1/2,1/2,0.297) develops below $T_N$.[13,14] The onset of order is sharp in temperature, although the magnitude of the heat-capacity step decreases with increasing x, and the residual heat capacity below the transition is low. The independence of $T_N$ on x is anomalous and is reminiscent of the pressure independence of $T_N$ in CeRhIn$_5$.[8] The relative insensitivity of $T_N$ to 'out-of-plane' doping contrasts to 'in-plane' doping with La in Ce$_{1-y}$La$_y$RhIn$_5$, where $T_N$ decreases smoothly with increasing x and vanishes for x~0.4.[15]

For 0.3 < x < 0.6, $T_N$ varies more dramatically with x, the magnetic transition broadens, and a second heat-capacity transition is observed in the vicinity of 1 K. The evolution of the magnetic transition in this regime is reminiscent of other doped heavy Fermion antiferromagnets such as CeRh$_{2-x}$Ru$_x$Si$_2$.[16] CeRh$_{0.5}$Ir$_{0.5}$In$_5$ is the most heavily studied representative of this doping range, and heat capacity, ac susceptibility and resistivity data are shown in Fig. 4. ac susceptibility measurements reveal that the 0.8-K transition in CeRh$_{0.5}$Ir$_{0.5}$In$_5$ is to a superconducting state (with 100% of full-shielding diamagnetism observed at the transition – as estimated based on measurements on a comparably sized and shaped sample of superconducting tin). The jump in heat capacity at the transition $\Delta C/\gamma T_c$ is of order 1, a value comparable to that observed in stoichiometric CeIrIn$_5$.[10] The temperature dependence of C below $T_c$ is also similar to that observed in CeIrIn$_5$,[17] suggesting that the nature of the unconventional superconducting gap is unchanged by the coexisting magnetic state. Preliminary muon spin rotation measurements on $CeRh_{0.5}Ir_{0.5}In_5$ reveal a static contribution to the magnetization developing below 3.8 K that is similar to that observed in $CeRhIn_5$ and persists to 100 mK.[18] These observations strongly suggest that superconductivity coexists microscopically with the ordered magnetic state. As will be discussed in more detail below, zero resistivity is observed in the vicinity of the diamagnetic transition but at a slightly higher temperature (for x=0.5, $T_{\rho=0}$=1K). In contrast to $CeIrIn_5$ in which there is no pronounced ac-susceptibility signature when the resistivity vanishes, a small shoulder is observed in $\chi_{ac}$ at $T_{\rho=0}$ in $CeRh_{0.5}Ir_{0.5}In_5$.

For x>0.6, ordered magnetism is completely suppressed and only a superconducting ground state is observed. Viewing this part of the phase diagram from the perspective of Rh-doping $CeIrIn_5$ reveals several remarkable features. At low Rh concentrations, $T_c$ appears to decrease with increasing Rh concentration, which is the conventional expectation, but then, in the vicinity of x ~ 0.9, $T_c$ recovers and increases with increasing Rh concentration. This feature in $T_c$ versus x for x~0.9 is somewhat reminiscent of the '1/8 anomaly' (i.e., 1-x ~ 1/8) observed in doped lanthanum cuprates[19] and is similar to what is observed for the pressure dependence of $T_c$ in superconducting $CeRhIn_5$.[20] In addition to the bulk phase transition observed in heat capacity data, a transition to zero resistance also persists over a wide range of doping. $T_{\rho=0}$ is relatively independent of doping so that $T_c$ approaches $T_{\rho=0}$ in the vicinity of x=0.5, resulting in a value of $T_c$ that is more than double that of stoichiometric $CeIrIn_5$. The upper critical field deduced from heat capacity measurements also increases significantly for x ~ 0.5 ($H_{c2}$=3 T compared with 0.5 T for $CeIrIn_5$), approaching values comparable to those required to produce a finite resistivity (5 T) in $CeIrIn_5$.[10] Finally, the anomalous shape of some of the heat capacity transitions in this concentration range (e.g., x=0.75 in Fig. 3) is reminiscent of early heat capacity data for $UPt_3$ and perhaps suggests the presence of a double superconducting transition.[21]

The remarkably rich phase diagram of $CeRh_{1-x}Ir_xIn_5$ reflects the inherent 'tunability' of ground states allowed by alternating $CeIn_3$:$MIn_2$ layers in $CeMIn_5$.[11] This behavior is not limited to $CeRh_{1-x}Ir_xIn_5$. Both $CeRh_{1-x}Co_xIn_5$ and $CeCo_{1-x}Ir_xIn_5$, although studied in less detail, show a similar multiplicity of phase transitions.[22] In the case of CeRhIn$_5$, the evolution from antiferromagnetic to superconducting ground state with pressure could be understood semi-quantitatively by considering the RhIn$_2$ layer as a source of chemical pressure on the CeIn$_3$ layer.[8] The similarity of the T-P phase diagram of CeRhIn$_5$ to the x dependence of T$_N$ in CeRh$_{1-x}$Ir$_x$In$_5$ suggests that the addition of Ir may act as an applied pressure. However, the in-plane lattice constant (which shrinks as a function of pressure) actually increases with increasing Ir concentration. Thus, 'hydrostatic chemical pressure' would appear to be an inadequate explanation of the observed similarities. On the other hand, 'uniaxial chemical pressure' appears to play a significant role in CeRh$_{1-x}$Ir$_x$In$_5$. If one considers the two ambient-pressure, stoichiometric superconductors in this family of compounds, CeIrIn$_5$[10] and CeCoIn$_5$,[11] one observes not only a substantial difference in T$_c$ (0.4 K vs. 2.3 K) but also a significant difference in *c/a*, the ratio of the tetragonal lattice constants (1.610 vs. 1.637). To the extent that a larger *c/a* implies greater anisotropy, larger T$_c$ for larger *c/a* is consistent with recent theories of magnetically mediated superconductivity.[23] The remarkable observation for the present CeRh$_{1-x}$Ir$_x$In$_5$ results is that the T$_c$ in CeRh$_{0.5}$Ir$_{0.5}$In$_5$ can be "predicted" based on the difference in *c/a* between CeRh$_{0.5}$Ir$_{0.5}$In$_5$ and CeIrIn$_5$ and the slope given by CeCoIn$_5$ and CeIrIn$_5$ (i.e., 0.8 K = 0.4K+0.006{1.9/0.027}).

The variation in *c/a* with Ir content, which to some extent must reflect changes in f-ligand hybridization, also produces characteristic trends in the magnetic susceptibility that correlate with the ground state configuration. Fig. 5 shows the anisotropic magnetic susceptibility for 2 K < T < 25 K for representative values of x in CeRh$_{1-x}$Ir$_x$In$_5$. $\chi_c$, the susceptibility for field applied along the *c*-axis, and $\chi_a$, the susceptibility for field applied along the *a*-axis, each reveal characteristic evolutions in T-dependence that coincide with different regions of the x-T phase diagram. With increasing x, the maximum in $\chi_c$ near 7 K, which can be associated with 2D magnetic fluctuations (and occurs above T$_N$),[8] is suppressed and evolves into a Curie-like increase. The maximum in $\chi_c$(T) disappears (i.e., $\chi_c$(7K) ~ $\chi_c$(2K)) for x~0.4, near the concentration at which superconductivity is first observed. The magnitude of $\chi_c$(7K) also reaches a maximum in this range. Thus, it appears that the spin fluctuation spectrum reflected in $\chi_c$ in these materials is coupled to superconductivity and is reminiscent of a broader trend observed in other CeMIn$_5$ compounds.[11] Accompanying these changes in $\chi_c$ is a corresponding evolution in $\chi_a(T)$: the loss of a maximum and subsequent drop in magnitude of susceptibility is centered around x~0.6, the concentration at which magnetic order is lost. The greater sensitivity of $\chi_a(T)$ to magnetic order is consistent with the *a*-axis being the easy magnetic direction. Thus, even at temperatures greater than the respective phase transitions, $\chi_c(T)$ signals the onset of superconductivity; whereas, $\chi_a(T)$ reveals the loss of magnetism.

As suggested above, substituting Ir for Rh must affect the degree of f-conduction electron hybridization. γ varies by a factor of two between CeRhIn$_5$ and CeIrIn$_5$ ($\gamma_{CeRhIn5}$=400 mJ/molK$^2$; $\gamma_{CeIrIn5}$=720 mJ/molK$^2$);[8,10] however, because the low-T heat capacity in CeRh$_{1-x}$Ir$_x$In$_5$ is dominated by magnetic fluctuations for low Ir concentrations, it is difficult to quantify the evolution of γ(x). Further, the complicated band structures for these materials make analysis of carrier density changes between them equally challenging, although their Fermi surfaces are rather similar.[24,25]

As discussed above, the coexistence of magnetism and superconductivity, rather than their competition, appears to be the rule and not the exception for U-based heavy Fermion superconductors. CeRh$_{1-x}$Ir$_x$In$_5$ clearly shares this feature. Preliminary photoemission data suggest another U-like characteristic of these materials.[26] Despite the observed large values of γ, no suggestion of a "Kondo resonance" near the Fermi surface is observed in these materials. Furthermore, local-density-approximation band structure calculations – which neglect many-body correlation effects – do a surprisingly good job of describing the electronic properties of CeMIn$_5$.[24,25] It would appear, then, that the relevant f-spectral weight is more band-like than localized, and perhaps such a situation is more susceptible to the coexistence of magnetism and superconductivity. Unfortunately, much of our present intuition derives from the localized limit.[27] Thus, the CeMIn$_5$ materials and CeRh$_{1-x}$Ir$_x$In$_5$ in particular may provide an opportunity to bridge our understanding of U-based and Ce-based heavy Fermion superconductivity.

In summary, we have presented a phase diagram for CeRh$_{1-x}$Ir$_x$In$_5$ that reveals several remarkable features. Superconductivity is observed over a very broad range of doping, 0.3 < x ≤ 1, including a substantial range of concentration (0.3 < x <0.6) over which it coexists with long-range magnetic order. The zero-resistance transition that is observed in stoichiometric CeIrIn$_5$ is robust against Rh substitution. In fact, the bulk T$_c$ more than doubles by CeRh$_{0.5}$Ir$_{0.5}$In$_5$, approaching the relatively unchanged temperature of the transition to zero resistance. The ground state evolution with Ir substitution appears to be 'controlled' by changes in c/a, reflective of variations in f-ligand hybridization.

We thank N.J. Curro for valuable discussions. Work at Los Alamos was performed under the auspices of the U.S. Department of Energy. Z.F. and P.G.P. acknowledge partial support from the N.S.F. (#DMR-9870034 and #DMR-9971348) and FAPESP (#9901062-0) respectively.

19. J.D. Axe et al., Phys. Rev. Lett. 62, 2751 (1989).

20. T. Muramatsu et al., unpublished.

21. G.R. Stewart et al., Phys. Rev. Lett. 52, 679 (1984).

22. P.G. Pagliuso et al., unpublished.

23. P. Monthoux and G.G. Lonzarich, to appear in Phys. Rev. B (2001).

24. Y. Haga et al., to appear in Phys. Rev. B (2001).

25. D. Hall et al., submitted to Phys. Rev. B (2000).

26. J.J. Joyce et al., unpublished.

27. S. Doniach, in Valence Instabilities and Related Narrow Band Phenomena, edited by R.D. Parks (Plenum, New York, 1977) p. 169.
Figure Captions

Fig. 1. Temperature-composition phase diagram of $CeRh_{1-x}Ir_xIn_5$. $T_{none}$ indicates the absence of (additional) phase transitions for $T \geq 350$ mK, the base temperature for our measurements. Lines are guides to the eye.

Fig. 2. Tetragonal lattice constants $c$ (circles) and $a$ (squares) for $CeRh_{1-x}Ir_xIn_5$. The solid line is a linear fit to $c/a$ (diamonds) as a function of x.

Fig. 3. Representative heat capacity divided by temperature (a) and magnetic entropy (b) versus temperature for $CeRh_{1-x}Ir_xIn_5$.

Fig. 4. Electrical resistivity, ac susceptibility, and heat capacity as a function of temperature for $CeRh_{0.5}Ir_{0.5}In_5$. The solid line is $C = 0.3\,T + 1.05\,T^2$, a temperature dependence similar to that observed in $CeIrIn_5$ [17].

Fig. 5. Anisotropic magnetic susceptibility for representative x in $CeRh_{1-x}Ir_xIn_5$. The upper cluster of curves are for H || c, and the lower cluster are for H || a. For a given concentration both susceptibility measurements were performed on the same crystal.

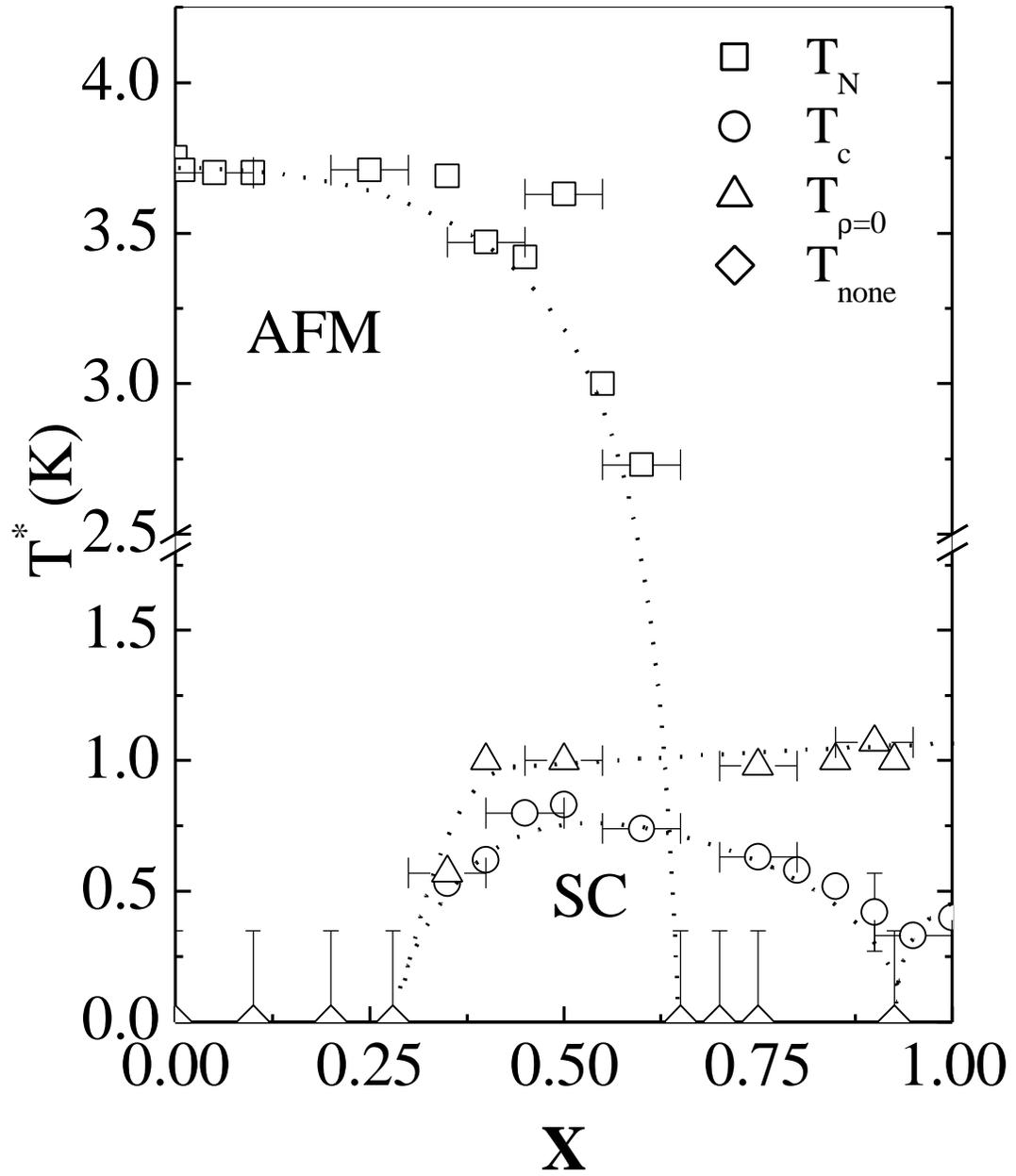

Figure 2, Pagliuso, *et al*

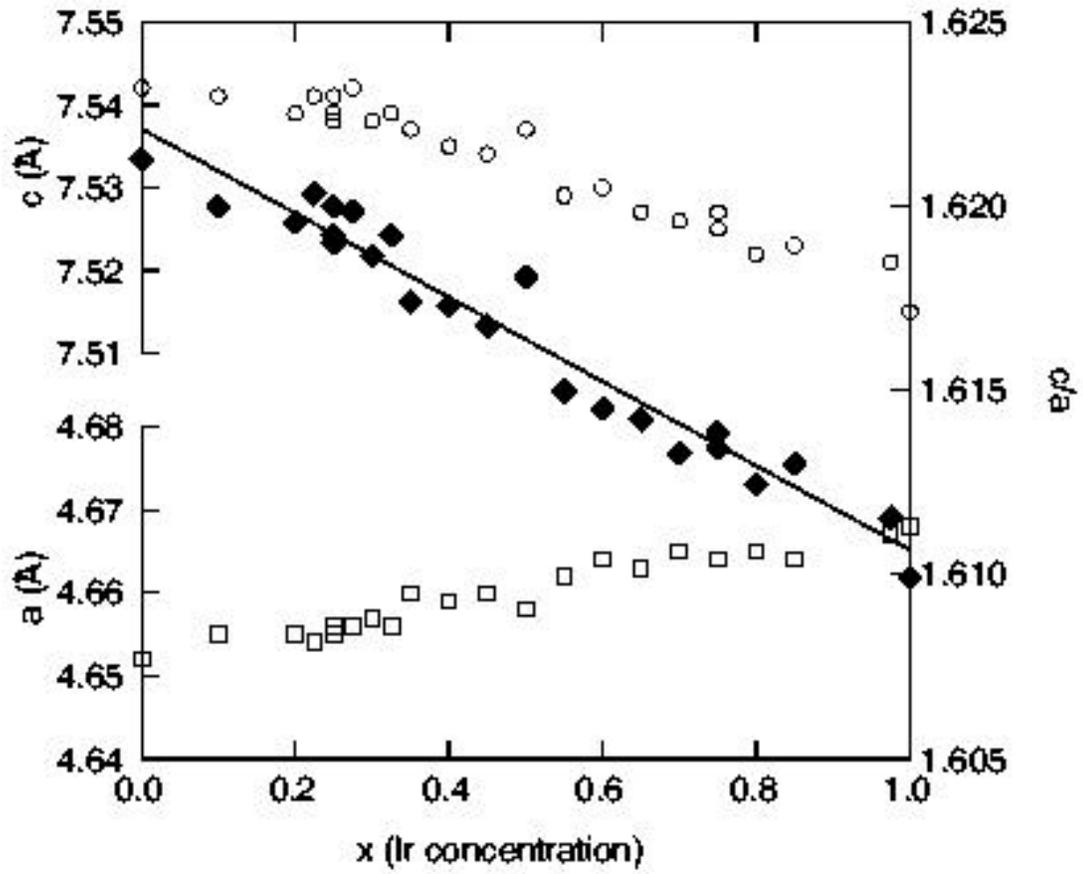

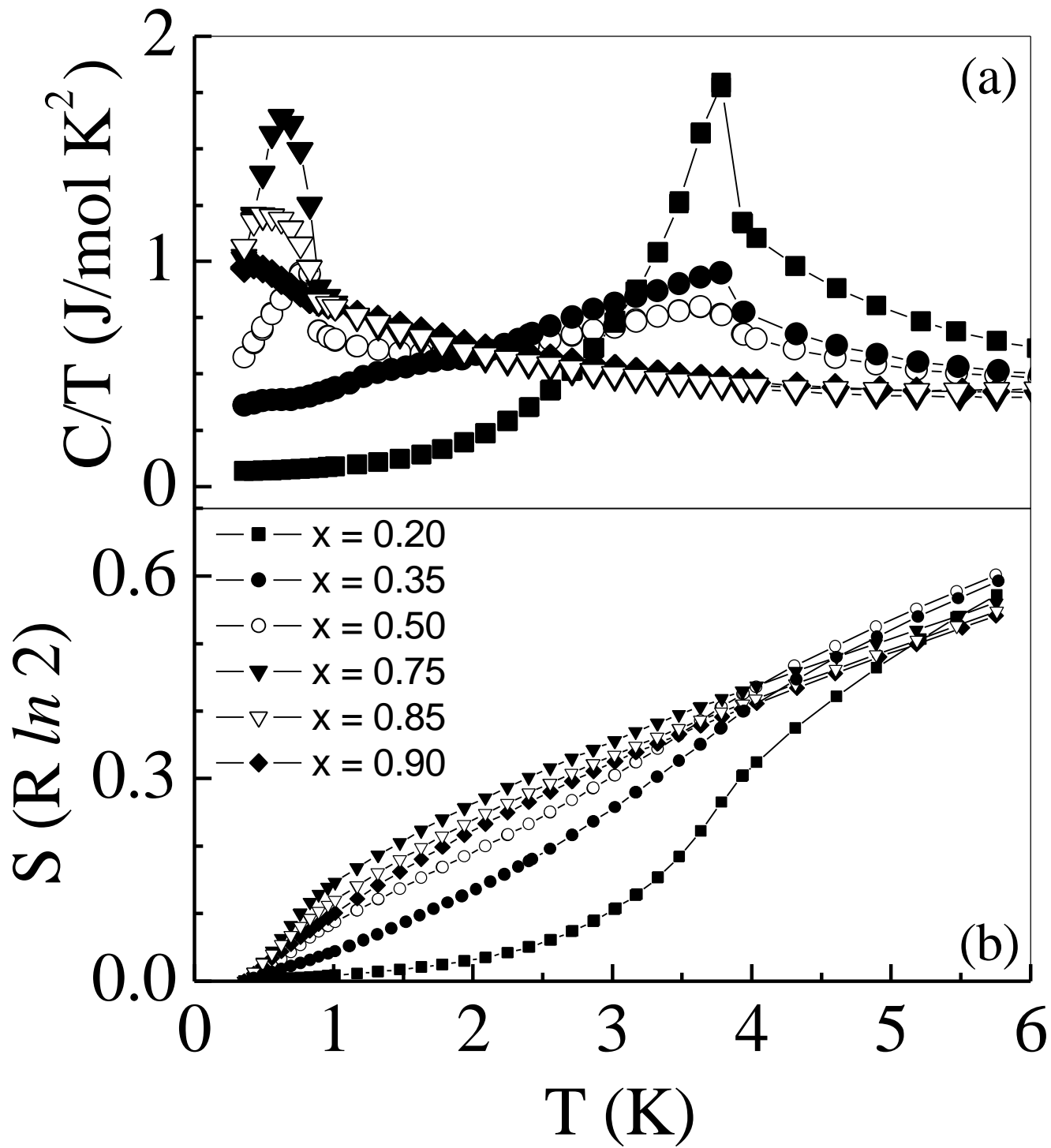

Figure 4, Pagliuso, *et al*

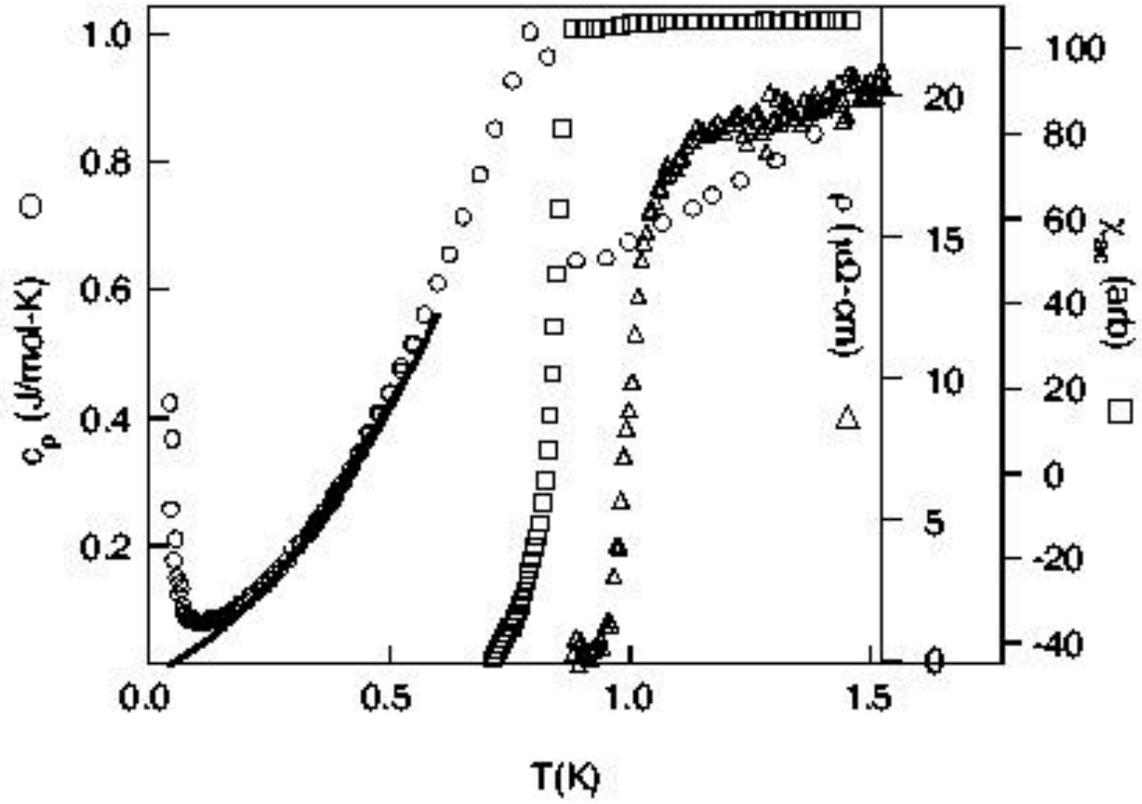

Figure 5, Pagliuso, *et al*

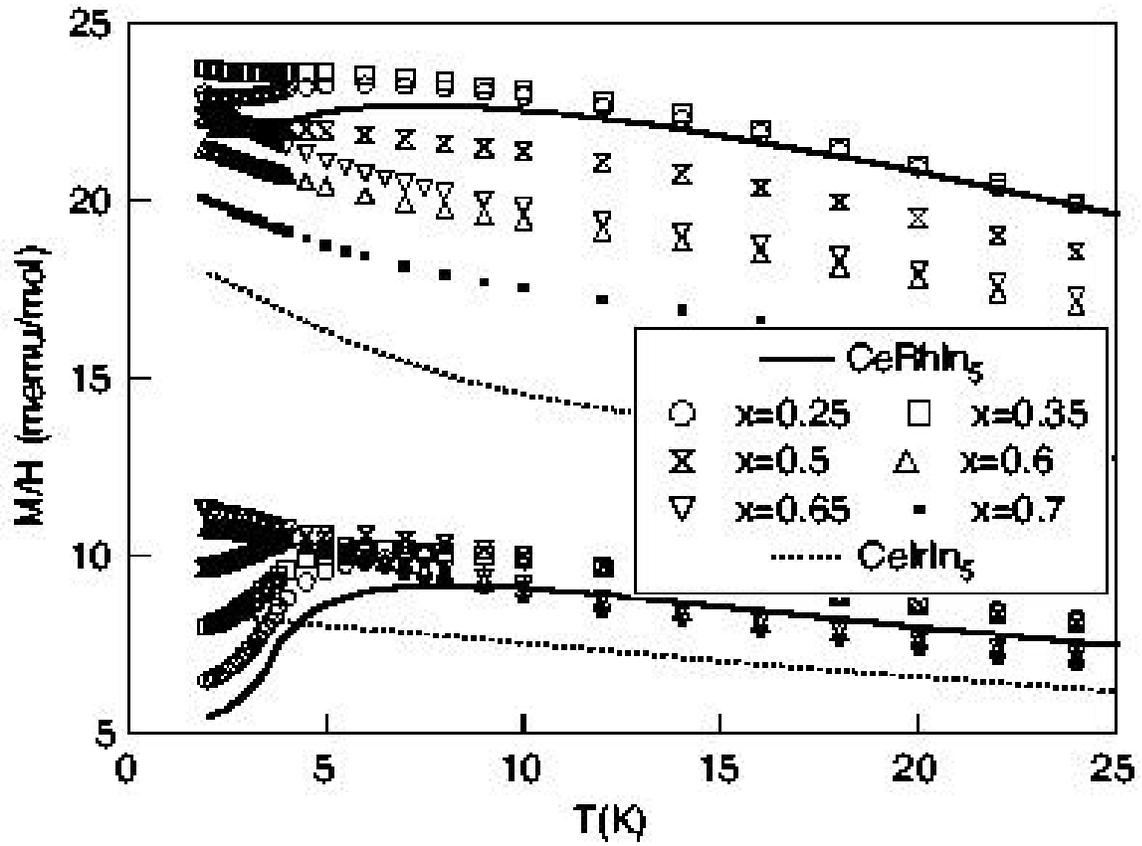